\documentclass[%
 preprint,
 amsmath,amssymb,
]{revtex4-1}

\usepackage{graphicx}
\usepackage{dcolumn}
\usepackage{bm}
\usepackage{hyperref}
\usepackage{amsmath}
\usepackage{multirow}
\usepackage{url}
\usepackage{subfigure}

\usepackage{caption}

\usepackage[utf8]{inputenc}
\usepackage{booktabs}
\usepackage{setspace}
\usepackage[table,xcdraw]{xcolor}

\usepackage{color}
\definecolor{redcolor}{rgb}{1.0,0.,0.}

\begin{document}

\preprint{}

\title{Using citation networks to evaluate the impact of text size on the identification of relevant concepts}

\author{Jorge A. V. Tohalino}%
\affiliation{%
Institute of Mathematics and Computer Science,  University of S\~{a}o Paulo,
S\~{a}o Carlos, SP,  Brazil
}
\author{Thiago C. Silva}%
\affiliation{Universidade Católica de Brasília, Brasília, DF, Brazil}
\author{Diego R. Amancio}%
\email{diego@icmc.usp.br}
\affiliation{%
Institute of Mathematics and Computer Science, Department of Computer Science, University of S\~{a}o Paulo,
S\~{a}o Carlos, SP,  Brazil
}%

\date{\today}

\begin{abstract}
The identification of the most significant concepts in unstructured data is of critical importance in various practical applications. Despite the large number of methods that have been put forth to extract the main topics of texts, a limited number of studies have probed the impact of the text size on the performance of keyword extraction (KE) methods. In this study, we adopted a network-based approach to evaluate whether keywords extracted from paper abstracts are compatible with keywords extracted from  full papers. We employed a community detection method to identify groups of related papers in citation networks. These paper clusters were then employed to extract keywords from abstracts. 
Our results indicate that while the various community detection methods employed in our KE approach yielded similar levels of accuracy, a correlation analysis revealed that these methods produced distinct keyword lists for each abstract. We also observed that all considered approaches, however, reach low values of accuracy. Surprisingly, text clustering approaches outperformed all citation-based methods. The findings suggest that using different sources of information to extract keywords can lead to significant differences in performance, and this effect can play an important role in applications relying upon the identification of relevant concepts.

\end{abstract}

\maketitle

\section{Introduction}\label{sec:intro}

With the increasing availability of large amounts of textual content on the Internet, the need for efficient analysis of texts has become imperative. Online textual data encompasses a wide range of sizes and types, including books, encyclopedias and  newspapers. In the last few decades, user-generated content in the form of short texts has also grown exponentially. Examples of such content include social media messages, product descriptions, online reviews, as well research papers~\cite{li2019key}.
In order to summarize this large amount of information, the task of keyword extraction (KE) has emerged as a crucial natural language processing (NLP) application. The goal of KE is to identify the most informative and relevant words or topics within a given document~\cite{timonen2012informativeness}.  Keywords serve as a useful tool for users, allowing them to quickly understand the overall content of the texts. Moreover, keyword extraction plays an important role in various NLP applications, including text categorization, document summarization, document tagging, recommendation systems, speech recognition, and many more~\cite{li2016textrank, li2019key}.

The KE task has been the subject of numerous studies. These investigations can be broadly classified into three categories: statistical methods, linguistic/syntactic approaches, and graph-based methods. Different paradigms have also been used in a combined approach for supervised classification, where the extracted features are employed in a machine learning algorithm for a binary classification task~\cite{jiang2009ranking}. While these methods have demonstrated effectiveness in processing large texts, they present significant challenges when applied to short texts with high sparsity~\cite{chen2020inside}. 

The identification of keywords within short texts, specifically in scientific manuscripts, poses a significant challenge, particularly when utilizing open scholarly datasets that only provide the title and abstract as sources of textual information~\cite{harzing2019two}. The challenge of extracting keywords from short texts, particularly in the case of scientific papers, has motivated the development of some approaches. One proposed method for addressing the sparsity of abstracts is to group abstracts using clustering techniques~\cite{silva2016using}. This can be accomplished by utilizing citations as a proxy for determining the similarity between papers, thereby circumventing the need for direct comparison of the short texts.
Despite the use of such clustering and other external information~\cite{li2019key,chen2020inside}, there is a lack of comprehensive studies comparing the compatibility of keywords extracted from abstracts and full texts. The accurate representation of the semantic information present in scientific papers is crucial in many areas, as it forms the foundation for many scientometric studies.
Thus, this study aims to address the following research questions:

\begin{enumerate}

\item To what extent are keywords extracted from abstracts similar to those extracted from the corresponding full papers?

\item Is there consistency in the set of keywords extracted by distinct community detection methods?

\item Does using citations result in superior performance as compared to directly assessing abstract similarity via textual information?

\end{enumerate}

We employed clustering methods to extract keywords from abstracts and compared them with keywords extracted from the corresponding full texts. Using a citation network, we evaluated the performance of various established community detection methods in identifying groups of related papers for the purpose of keyword extraction. We  also evaluated clustering approaches that do not rely on citation information, including techniques based on neural embeddings.

The study revealed several interesting results. All evaluated methods were found to have a considerable discrepancy with keywords found in the full texts. 
We observed that clustering methods that rely solely on textual information outperformed those based on citation networks, indicating that citations may not be an optimal proxy for semantic similarity. Furthermore, our results indicate that the various community detection strategies evaluated yielded similar performance levels, despite the observed differences in the set of keywords identified by each approach.

In summary, our findings suggest that the quantity of information used to extract keywords can strongly impact the performance of the task. Therefore, studies using similarity networks should consider the use of full texts, when available, to provide more robust information regarding topics extracted from paper networks.

The structure of this paper is as follows: In Section~\ref{sec:related}, we present a comprehensive review of the most pertinent studies in the area of keyword extraction. The proposed methodology for extracting keywords from both short and long texts is outlined in Section~\ref{sec:methods}, which also includes information regarding the adopted datasets. The main results are presented and discussed in Section~\ref{sec:results}. Finally, in Section~\ref{sec:conclusion}, we summarize with  conclusions and suggest potential perspectives for future research.

\section{Related works}\label{sec:related}

The early works that addressed the keyword extraction problem focused on statistical methods.  The spatial distribution of words along the text is used to gauge words' relevance~\cite{carretero2013improving}. The most simple approach is based on word frequency, where words with higher frequency values are considered keywords. However, these methods do not consider word order, therefore, if the text is shuffled, a meaningless version of the text would generate the same set of keywords. The combination of frequency and spatial distribution was then proposed to address this issue via word clustering and  entropy~\cite{carretero2013improving,ortuno2002keyword}. The idea behind these methods is that important words are commonly concentrated in certain parts of the text, where the main topics are located. In this sense, irrelevant words are distributed regularly along texts, while keywords present an uneven distribution and tend to form semantic groups. 
Another improvement to frequency-based methods is the tf-idf approach, which weights the importance of a word according to its frequency within a text and the frequency along the dataset. The main advantage of these methods is that they are simple and do not require an external corpus or knowledge of the language.

Graph-based methods have also been used approaches to model texts~\cite{machicao2018authorship}. Several works addressed the keyword extraction problem representing documents as word co-occurrence networks, where two words are connected if they co-occur in a given context~\cite{tohalino2022using}. 
Centrality metrics are then used to assign an importance value to each word. 
In~\cite{lahiri}, the authors concluded that network metrics are able to successfully extract relevant words for the keyword extraction task. They also highlighted that network-based approaches do not need the use of external corpora and they are language independent. The use of word embeddings and large contexts has also been useful in improving the quality of co-occurrence networks when extracting keywords~\cite{tohalino2022using}. A different approach was proposed by~\cite{grineva2009extracting}, where community detection methods were applied to a network of semantic relationships. The authors used Wikipedia to establish the semantic relatedness between the words of the document. According to~\cite{grineva2009extracting}, important words tend to be grouped into highly connected communities, which are related to the main topics of the document.       

In order to address the keyword extraction problem in short texts, several works rely on the use of semantics and background knowledge. According to to~\cite{chen2020inside}, extracting only basic or straightforward features from the words is insufficient for finding keywords from short texts. In \cite{li2019key}, the authors remarked that text clustering approaches could also be useful in addressing the semantic sparseness of short texts. 
These techniques enable the clustering of related texts, thereby allowing for the extraction of more semantic information by aggregating texts in the same cluster. 
\cite{wan2008collabrank} employed clustering algorithms to identify the most relevant words for each cluster, operating under the hypothesis that texts with similar topics contain similar keywords. Then, a graph-based approach was applied to each text cluster; and the PageRank algorithm was used to extract keywords.

In~\cite{timonen2012informativeness}, the authors proposed a method for selecting keywords based on the informativeness value of each word.
This score was calculated at the corpus, cluster, and document levels. At the corpus level, the informativeness was computed taking into account all the documents, while at the cluster level, the word importance was calculated within a group of related texts. The results from the previous steps were then used to compute the informativeness at the document level. This approach yielded a good performance for extracting keywords.

Regarding graph-based techniques, several studies have enhanced  TextRank~\cite{mihalcea2004textrank} by incorporating different semantic relationships between words as node weights for the word ranking algorithm. For instance, \cite{li2016textrank} used Wikipedia as an external knowledge base, while~\cite{li2019key} employed the Word2Vec and Doc2Vec embedding models to compute the semantic similarity between words.

While many works focus on extracting keywords either from short- and long-texts, here we conduct a comparative analysis of well-established methods for extracting keywords from both short and long texts. We focus on determining the compatibility of keywords extracted from abstracts and those extracted from the full content of research papers.

\section{Material and methods}\label{sec:methods}

\begin{figure}[h]
    \centering
    \includegraphics[width=1.0\textwidth]{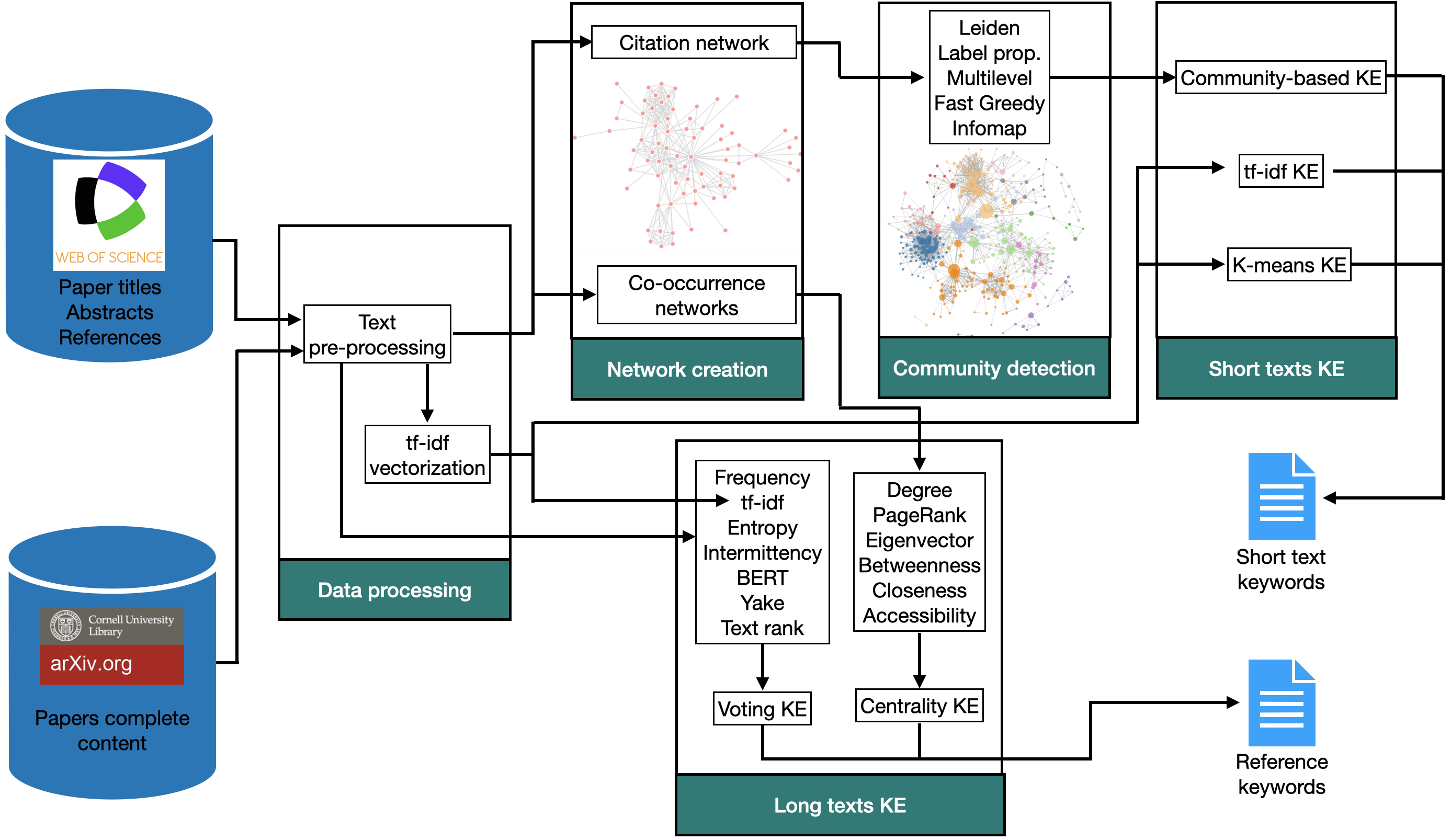}
    \caption{The workflow of the adopted keyword extraction system involves several stages. The initial step is pre-processing of the text. Next, a citation network is constructed and community detection algorithms are employed to extract keywords from short texts. For comparative evaluation, we implemented both a tf-idf and K-Means approach. Both approaches do not consider citations to cluster papers.
 In order to generate reference keywords, we employed a combination of statistical and traditional keyword extraction methods on the full texts. Additionally, we also evaluated a co-occurrence network approach as an alternative method for keyword extraction.
}
   \label{fig:architecture}
\end{figure}

The framework proposed to extract keywords 
comprises the following main steps: i) text pre-processing; ii) network construction; iii) community detection; iv) short texts keyword extraction; and v) long texts keyword extraction.  The steps are summarized below and illustrated in Figure \ref{fig:architecture}.

\begin{enumerate}

    \item \emph{Text pre-processing}:  this phase comprises the text-processing and vectorization steps. 
    The first step includes the removal of stopwords. The remaining words are stemmed and the tf-idf approach is employed to obtain the vectorized form of the pre-processed texts. Additional details on the pre-processing steps applied can be found in Section~\ref{sec:processing}.
    
    \item \emph{Network creation}: we first constructed a paper citation network, which is used for short text keyword extraction. We also modeled the complete content of each paper as word co-occurrence networks, which were used to extract keywords from long texts. In Section~\ref{sec:network}, we describe the required steps for the creation of both network models.  
    
    \item \emph{Community detection}: we applied community detection methods to the citation networks in order to find clusters of related papers (see Section \ref{sec:community}).
    
    \item \emph{Short texts keyword extraction}: this phase is responsible for the extraction of keywords from short texts (paper abstracts). The clusters obtained in the previous step are used in this phase. The relevance of each word is computed inside and outside communities. We also proposed two methods for keyword extraction based on tf-idf and the K-Means algorithm. The methods for short texts keyword extraction are described in Section ~\ref{sec:short}.   
    
    \item \emph{Long texts keyword extraction}: to identify reference keywords, we used the complete content of the papers as input from several well-known keyword extraction methods. We evaluated methods based on word frequency, tf-idf, entropy, intermittency,  BERT, Yake and TextRank~\cite{carretero2013improving,amancio2013probing,campos2020yake,mihalcea2004textrank,devlin2018bert}. We also used a network approach based on co-occurrence networks and centrality metrics to find keywords for long texts. These networks were characterized using centrality metrics. A detailed explanation of the adopted methodology is shown in Section~\ref{sec:long}.   
                
\end{enumerate}

\subsection{Datasets}\label{sec:dataset}

The following two datasets were used:

\begin{enumerate}
    
    \item \emph{Short texts KE dataset}: we used the dataset collected in~\cite{silva2016using}. The authors retrieved the information from 11,063 papers on the complex networks field. The data was obtained from the \emph{Web of Science} (WOS) database~\footnote{https://clarivate.com/webofsciencegroup/solutions/web-of-science/}. The selected papers were published from 1991 to 2013. For each paper, the authors extracted the title, abstract, and list of references. The latter was used to construct a citation network. The title and abstract of each paper were used as input for the application of keyword extraction techniques.
    
    \item \emph{Long texts KE dataset}: In order to generate a list of reference keywords for each abstract, we collected the full content of each paper, including the introduction, methodology, results and discussion, conclusions, and appendix sections of each research article. We used the API of the arXiv database \footnote{https://arxiv.org/} to extract the complete content of the papers. We performed an automatic search using both the title and the abstract of each paper as keywords for the arXiv API.

\end{enumerate}

Table~\ref{tab:dataset} presents a summary of the statistical information for the datasets. The information provided was calculated from the pre-processed versions.  

\begin{table}[h]
\caption{Statistical information from datasets. $|D|$ stands for the number of documents. We also show the average number of tokens ($W_{avg}$), and the first ($W_{q1}$) and third ($W_{q3}$) quartiles of the distribution of the number of tokens. Similarly, $U_{avg}$ represents the average vocabulary size, while $U_{q1}$, and $U_{q3}$ are the first and third quartiles of the vocabulary size distribution, respectively.} 
\label{tab:dataset}
\begin{tabular}{c|c|c|c|c|c|c|c|c}
\hline \hline
\textbf{Dataset} & \textbf{Description} & $|D|$ & $W_{avg}$ & $W_{q1}$ & $W_{q3}$ & $U_{avg}$ & $U_{q1}$ & $U_{q3}$ \\ \hline
Short texts & Paper abstracts  & 11,063 & 79.12 & 59.00 & 95.00 & 55.97 & 43.00 & 66.00 \\ 
Long texts & Full paper content  & 1,982 & 2,020.58 & 1,170.25 & 2,405.00 & 517.64 & 394.00 & 595.00 \\ \hline \hline
\end{tabular}
\end{table}

\subsection{Data processing}\label{sec:processing}

This phase comprises three steps: data preparation, text pre-processing, and tf-idf vectorization. The data preparation step consisted of processing the recovered papers from the arXiv database (for the full content of the papers). We obtained a LaTeX version of each paper, so we had to remove all LaTeX tags. We also removed  the authors list, institutions, and acknowledgments from the cleaned text. The following sections were included in the analysis of full papers: introduction, related works, methodology (or materials and methods), results, discussion, conclusion, and appendix sections.

Text pre-processing transformations were applied to all texts of the dataset. We first removed stopwords and punctuation marks. Then, a stemming step was applied to the remaining words. This step is required in order to map each word into its root or stem~\cite{pramana2022systematic}. 
The tf-idf technique was used to transform the pre-processed text into a sparse vector representation. To compute the importance of a word $w$, the technique considers the internal frequency of $w$ in a single document. Moreover, the internal frequency is compared with  the relative frequency of $w$ in all documents of the dataset~\cite{salton1973specification}. 
The tf-idf representation of $w$ in a document $d$ is computed as 
\begin{equation}
    \textrm{tf-idf}(w,d) = \frac{f(w,d)}{n_d} . \frac{\log N}{\log (N_w)}, 
\end{equation}
where $f(w,d)$ represents the frequency  $w$ in $d$, $n_d$ is the number of words in  $d$, $N$ stands for the total number of documents in the dataset, and $N_w$ represents the number of documents in which $w$ appears at least once.

We used the tf-idf vector representations of each abstract as input values of a K-Means based method for short texts KE. We also used the tf-idf weight of each word from the full content of the papers in order to give an importance value  for a long text KE method (LKE). 

\subsection{Network creation}\label{sec:network}

Given the short size of paper abstracts, 
it is infeasible to extract statistically significant information from individual texts. As such, we employed network representation techniques to extract supplementary information that would enhance the keyword extraction process. Upon analyzing the topological and structural properties of the networks, we are able to infer attributes of the texts that enable us to determine the relative importance of each word.

Two different network models are used in our study. 
In order to cluster \emph{short texts} into groups of related papers, we used a citation network for the \emph{short texts} keyword extraction task. In this case, the network structure represent the whole dataset of documents. Conversely, when extracting keywords from long texts,
each text is modeled as a word co-occurrence network~\cite{castro2019multiplex,joseph2022cognitive,stella2020multiplex}.

The unweighted paper citation network was built following the methodology described in~\cite{silva2016using}.
The citation networks are intended to represent a semantical similarity structure that do not use textual information to establish links between papers. The resulting network was composed of $11,063$ nodes and $94,472$ edges. The community structure of this network and the information of title and abstract are then used to detect the most important words in each network community. 

The \emph{full content} of a paper is modeled as a word co-occurrence network. In this graph model, each node represents a word, and the edges between two nodes are based on the neighborhood relationship of two words. We used the approach that can include virtual links, so that similar words can be linked. This model and its variations have been used in many different scenarios~\cite{de2016topic,correa2019word,ferraz2018representation,quispe2021using}. The networks were characterized using well-known centrality measurements to rank the words according to their structural importance in the networks~\cite{newman2018networks}. 

\subsection{Community detection} \label{sec:community}

This phase is responsible for detecting communities, i.e. clusters of papers linked via citation links. Communities are groups of nodes that are more densely interconnected with each other in comparison with the rest of the nodes from the network~\cite{radicchi2004defining}. The identification of communities in large networks is quite a useful task. For example, the nodes that belong to the same community likely share several common properties. Also, the number of found communities and their respective features could help to identify the category of a network for classification tasks~\cite{wang2015network}. The identification of communities is also useful to understand the dynamic evolution and organization of a network~\cite{dakiche2019tracking}. 
In this paper, we evaluated the following methods: Multilevel, Label Propagation, Infomap, Fast Greedy, and Leiden method~\cite{Blondel_2008,PhysRevE.76.036106,traag2019louvain,clauset2004finding,rosvall2008maps}. In the Appendix, we provide a brief description of each method.

In the context of community detection methods, we investigated if community-based methods are consistent in the sense that they generate well-defined, large communities. This is an important step in our analysis because small communities can lead to low performance~\cite{silva2016using}. In the paper citation network, most of the community detection methods found between 23 and 39 paper communities, which leads to communities comprising more than 100 papers, typically. The infomap, however, generated more than 400 communities, and most of them comprised less than 10 papers. {Before the computation of the relevance of each word, we decided to filter out those communities that contain few papers. }

\subsection{Short texts keyword extraction}\label{sec:short}

This step consists of the extraction of keywords from the pre-processed paper abstracts. We evaluated a network community-based approach that generates a word importance index to rank each word from the paper abstracts. For comparison purposes, we also evaluated tf-idf and K-Means-based methods for the short texts KE task. 

\begin{enumerate}

    \item \textit{Community-based approach:} we used the community structure found from papers citation networks to detect the word importance index of each word from paper abstracts. The adopted index quantifies the relative frequency of a word appearing inside a community against its frequency in the remaining documents of the citation network~\cite{silva2016using}.  To compute the word importance index $I$ for a word $w$, we first compute the frequency of the word inside a community $\alpha$. This quantity is the relative internal frequency $F_\alpha^{\textrm{(in)}}(w)$, given by
    \begin{equation}
        F_\alpha^{\textrm{(in)}}(w) = \frac{n_\alpha(w)}{|\alpha|},
    \end{equation}
    where $n_\alpha(w)$ is the total number of papers containing $w$ appears within a community $\alpha$, and $|\alpha|$ represents the number of papers associated with a community $\alpha$. We also compute the relative frequency of $w$ outside $\alpha$, $F_\alpha^{\textrm{(out)}}(w)$, which is computed as:      
    \begin{equation}
        F_\alpha^{\textrm{(out)}}(w) = \sum_{\gamma \neq \alpha} \frac{n_\gamma(w)}{N-|\alpha|},
    \end{equation}
    where $N$  is the total number of papers in the network. Then, the importance index $I(w)$ is calculated as the highest difference between the relative in-community and out-community frequencies, i.e.:
    \begin{equation}
        I(w) = \max_\alpha \Big{[} F_\alpha^{\textrm{(in)}}(w) - F_\alpha^{\textrm{(out)}}(w) \Big{]}.
    \end{equation}

    The word importance index was computed for all words from paper abstracts, and then the best-ranked words were considered as relevant keywords for each abstract.
   
    \item \textit{tf-idf based approach:} the tf-idf values considering all paper abstracts from the dataset are computed. For each abstract, we considered the tf-idf weights of the words comprising the abstract. The words with the highest tf-idf values were selected as relevant keywords for each abstract.  

    \item \textit{K-Means based approach:} this method is equivalent to the \emph{community-based approach}. The difference is that clusters are obtained via the K-Means algorithm~\cite{rodriguez2019clustering}. To obtain the cluster, we first obtained the embedding of each abstract. Then we evaluated several values of $K$ to find the optimal number of clusters.
    
\end{enumerate}

\subsection{Long texts keyword extraction} \label{sec:long}

The keywords obtained from full texts are considered reference keywords when evaluating the quality of keywords extracted from short texts. 
%
Here we considered as input texts the complete content of the research papers. We adopted several methods found in the literature to extract keywords documents. 
The methods can be classified into two approaches: statistical and graph-based approaches:
   
\begin{itemize}

    \item \textit{Statistical and traditional keyword extraction methods:} In this step we employed statistical techniques that are commonly used for keyword extraction tasks. These methods perform an appropriate analysis of the statistical distribution of words along documents. The main goal of statistical methods is to detect and rank relevant words of documents without any \emph{a priori} or external information~\cite{carretero2013improving}. The methods we adopted are based on frequency, word tf-idf, word entropy, word intermittency, and Yake. We also evaluated a graph-based approach named TextRank, and a method that uses word embeddings based on BERT. The methods are described in the Appendix.
    
    \item \textit{Network-based methods:} 
    a comprehensive set of centrality measures were used to analyze the word co-occurrence networks derived from the full content of the papers.
    The network measurements are useful to identify the most relevant nodes in a network~\cite{newman2018networks}. Therefore, they allow ranking the nodes according to their topological importance so that they can find the most important words for each text~\cite{tohalino2018extractive}. We selected as keywords for each text the best-ranked nodes (words) according to the following centrality metrics: degree, PageRank, betweenness, eigenvector centrality, closeness and accessibility computed at the first two levels~\cite{travenccolo2008accessibility}. We also employed a methodology that combines the results of each centrality metric. In the methodology henceforth referred to as voting system, the keywords found by the majority of the network measurements were selected as relevant keywords for each text.
    
\end{itemize}

\section{Results and discussion}\label{sec:results}

%
Our analysis is divided into two sections. Section~\ref{sec:dataset_analysis}  describes a statistical analysis of the datasets.   Section~\ref{sec:accuracy_analysis} provides a comparison of keywords extracted from short and full-text sources. We also analyze the performance of distinct network community methods for the task. 

\subsection{Dataset analysis and selection of reference keywords}\label{sec:dataset_analysis}

In this section, we first perform a statistical evaluation of the datasets through the analysis of the number of common words between the paper abstracts (short-size texts) and the full content of the research paper (long-size texts). This analysis is an initial step to the generation of a set of gold standard keywords for each paper abstract. Because many datasets comprising full-text papers lack keywords selected by human experts, we used as a starting point the full content (including all sections except the abstract) of each paper.  We employed keyword extraction methods to extract reference keywords from the complete content of each paper. However, we first analyzed the number of mutual words existing between each abstract and the full content. In some cases, it is possible that the paper authors use specific words to express their main ideas in the abstract and they could change to other words using synonyms or similar expressions in the rest of the paper. Therefore, it becomes important to analyze whether the information extracted from full texts is compatible with the content of abstracts.

\begin{figure}[h]
    \centering
    \includegraphics[width=0.9\textwidth]{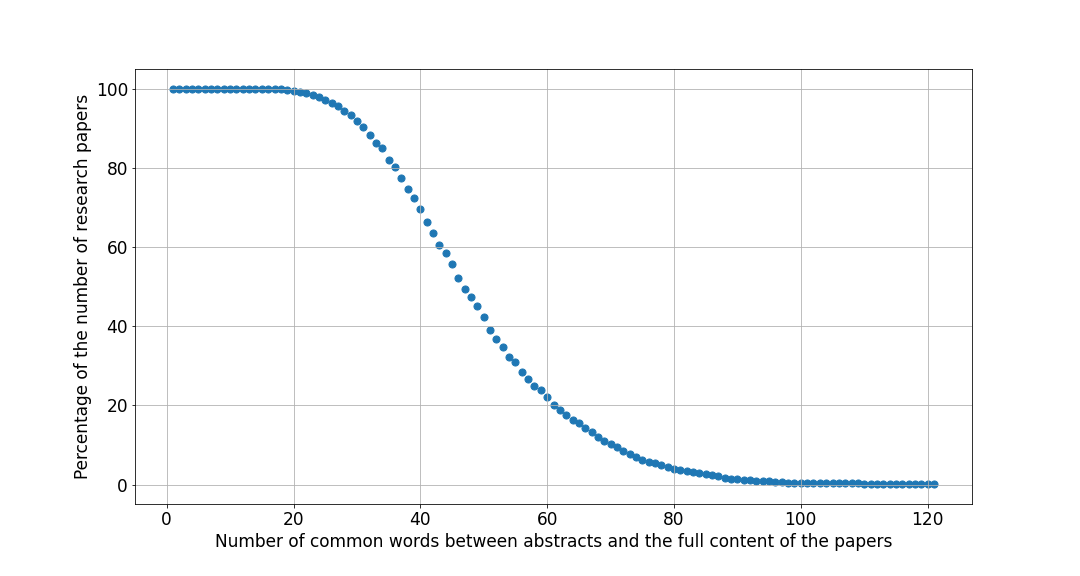}
    \caption{Analysis of the number of common words between the abstracts and the complete content of the research papers. The x-axis represents the number of mutual words between paper abstracts and their corresponding full content. The y-axis is the representation of $P(x \geq w)$, i.e. the fraction of papers comprising at least $w$ common words between abstracts and full texts.
    }
   \label{fig:common_words}
\end{figure}

We computed how many words ($w$) in the abstract are also present in the full content of the papers. The cumulative distribution (i.e. $P (x \geq w$)) of this quantity in the dataset is shown in Figure~\ref{fig:common_words}. 
A significant number of research papers (80\%) present a high number of common words (40) between the abstracts and the full content of the papers. 50\% of the papers have at least 50 common words.
As expected, this means that most of the information in the abstract is also available in the remainder of the paper.

\begin{figure}[h]
    \centering
    \subfigure[Analysis of traditional and statistical keyword extraction methods for the full content of research papers.]{\includegraphics[width=0.8\textwidth]{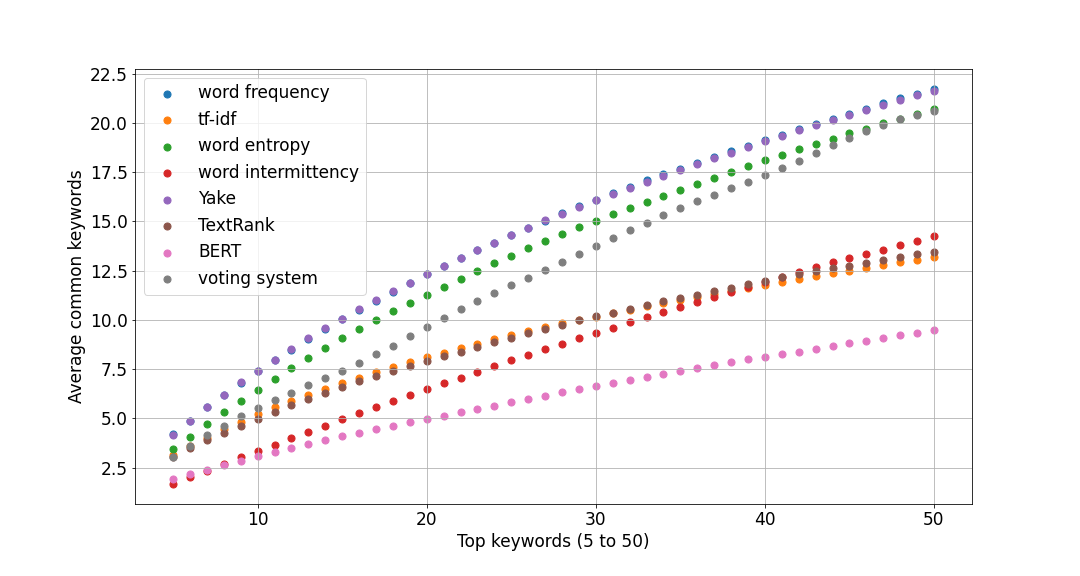}\label{fig:common_words2_a}} 
    \subfigure[Analysis of network-based keyword extraction methods for the full content of research papers.]{\includegraphics[width=0.8\textwidth]{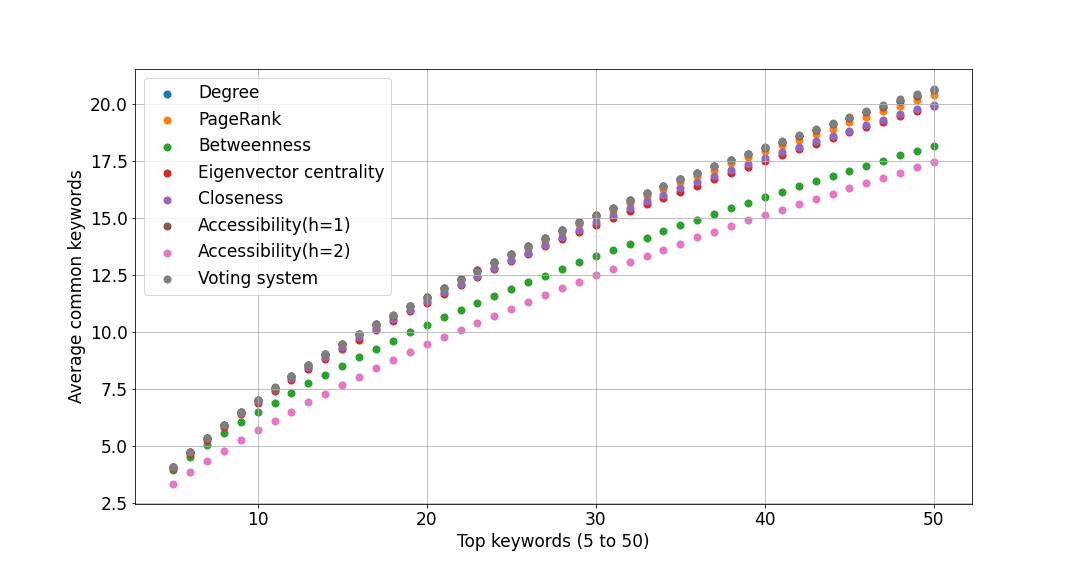}\label{fig:common_words2_b}} 
    \caption{Analysis of the overlap between the common words found in the paper abstracts and those identified by keyword extraction methods for longer texts (i.e., the full content of the papers).
    The x-axis represents the number of keywords recovered in the full content, while the y-axis indicates the average number of retrieved keywords that also appear in the paper abstract.  
    }
    \label{fig:common_words2}
\end{figure}

Now we evaluate how many \emph{keywords} found in the \emph{full content analysis} are also present in the abstract.
We used two approaches to extract reference keywords considering each paper's full content: statistical and graph-based KE methods. We evaluated these approaches by counting the number of mutual words between the keywords generated by each method and the words composing the paper abstracts. Figure~\ref{fig:common_words2} depicts the obtained results for each approach.  
According to the size of the abstracts and the full content (see Table~\ref{tab:dataset}), we considered recovering between 5 and 50 keywords generated by each KE method. Then, we count the number of these keywords that are part of the abstracts.  

In relation to traditional and statistical methods, the results displayed in Figure~\ref{fig:common_words2_a} show that the methods Yake, word frequency, and word entropy outperformed the other KE techniques. These methods were able to find the largest number of common words with the abstracts. 
The voting system approach did not achieve the best results. The methods based on word intermittency and BERT also displayed a low number of mutual words with the paper abstracts. 
We also evaluated the methods based on word co-occurrence networks and centrality measurements. The results depicted in Figure~\ref{fig:common_words2_b} suggest that almost all centrality metrics performed similarly. We observed that the voting system, node degree, PageRank, and accessibility (computed at the first hierarchical level) outperformed the other network-based methods. However, the difference in terms of performance with the other network metrics is not significant.

\subsection{Extracting keywords from abstracts} \label{sec:accuracy_analysis}

In this section, we analyze if the methods adopted to extract \emph{keywords from abstracts} are able to capture keywords that are found when the full-text content is analyzed. We used accuracy as a performance evaluation measure.
The performance of the methods is measured in terms of the number of common words between the reference keywords and the keywords generated by the short-text KE methods, divided by the total number of reference keywords. 
We established a parameter $N$ to represent the number of reference keywords to be considered in the evaluation. As reference keywords (i.e. keywords obtained from full texts), we used the methods with the highest performance observed in the previous section.

\begin{figure}[h]
    \centering
    \subfigure[~ Performance analysis considering the traditional and statistical methods as reference keywords. 
    ] {\includegraphics[width=0.8\textwidth]{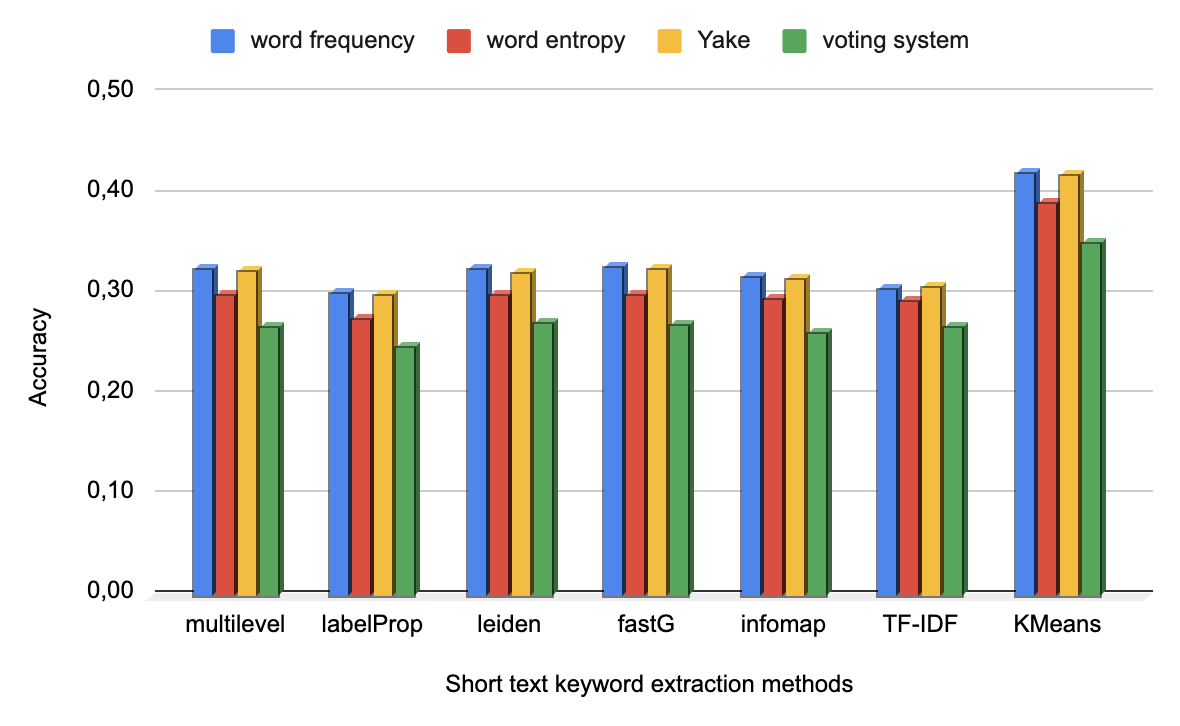}\label{fig:accuracy_a}}
    \subfigure[~ Accuracy analysis considering the network-based methods as reference keywords.] {\includegraphics[width=0.8\textwidth]{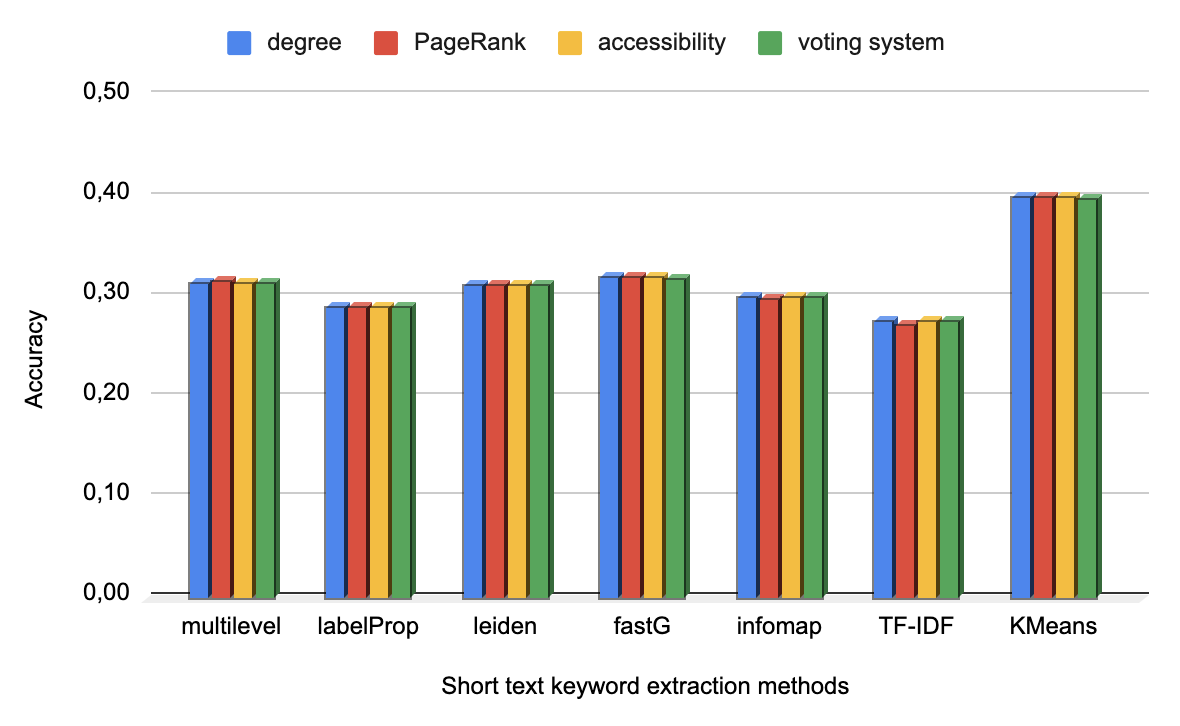}\label{fig:accuracy_b}} 
    \caption{ Comparative analysis of the accuracy obtained from the evaluation of KE methods for short texts (paper abstracts). We extracted $N=30$ keywords  from full texts.}
    \label{fig:accuracy}
\end{figure}

Figure~\ref{fig:accuracy_a} shows the performance analysis considering  as reference keywords the ones obtained from statistical methods. 
The results in Figure~\ref{fig:accuracy_a} suggest that community-based approaches obtained similar accuracy values since no method clearly outperformed the others. The label propagation method achieved a slightly lower performance than the other methods. We also found that the tf-idf method displayed a performance that is similar to the other network community-based methods. Surprisingly,  when citation information is disregarded and only the textual information is used, the performance is improved. The K-Means method is significantly better than all other approaches, with a gain of 25\% in performance, in some cases. {The complete analysis considering different values for the parameter $N$ is shown in the Appendix.} 

Figure~\ref{fig:accuracy_b} shows the performance analysis considering as reference keywords the ones obtained from (co-occurrence) network-based methods. Here we see that all considered network centrality metrics provide almost the same performance. This result is consistent with the literature in text network analysis since there is a correlation of centrality metrics when analyzing written texts. Concerning the formation of reference keywords via community detection methods, it is also worth noting that  all community-based methods provided similar performance for the task. Conversely, choosing reference keywords via tf-idf yielded the worst performance.  Once again, the best performance was obtained with the KMeans method.

One possible explanation for the similar performance achieved by the community-based detection methods could be the fact that all methods are generating similar partitions and, consequently, they are selecting the same set of keywords for each paper abstract. In order to evaluate this hypothesis, we computed the Spearman correlation coefficient of the ranking of words' relevance generated by different community detection methods. The results (not shown) revealed that the methods are actually selecting different sets of keywords. If one considers the full rank of words, the Spearman correlations are typically lower than 0.10. In a similar fashion, when considering only the top 30 ranked words, all correlations were below 0.27. As expected, in both scenarios,  the highest correlation was found for the rankings generated by the Multilevel and Leiden methods~\cite{traag2019louvain,Blondel_2008}.

The performance results revealed interesting insights. First, we found that identifying keywords from citation information alone does not provide the highest match between keywords found in abstract and full-text. While citations have been used in numerous contexts~\cite{amancio2012using}, one possible reason for the observed low performance is that citations may not reflect the semantic similarity of texts, which may hinder the performance of the community detection methods. In fact, some studies have pointed out a discrepancy between citation and content similarity. For example,~\cite{amancio2012using} found that citation and content similarity are not consistent  since the most similar papers are oftentimes disregarded when selecting references for papers. In a similar fashion, the differences in the content have been used to improve models reproducing the growth of citation networks, since content similarity has also been used as an important feature to model the growth of citation networks~\cite{zhao2022utilizing}.

While the use of textual information was able to provide a better performance in recovering keywords from full-text content, the obtained accuracy is still below $50\%$. This means that using cluster information from papers abstracts is not enough to recover the full content of papers. This may have implications in many studies that are based on recovering text content based on keywords. For example, when studying the properties of citation networks, the selection of papers via keywords may affect the stability of citation network metrics. A different number of communities depicting subfields of a major area can be found if the keywords terms are not well-defined to select the relevant papers. 

The differences in content extracted from abstracts and full texts can also potentially lead to distinct interpretations in the context of \emph{Science of Science}. In a document similarity network, for example, the centrality of a paper may strongly depend on the use of abstracts or full texts. If such networks are studied in other contexts, this may lead to less robust conclusions. For example, comparing the semantic similarity between papers linked by citations may lead to different results depending on how much text is used to gauge semantic similarity~\cite{amancio2012using}.  
Therefore, it remains relevant to consider full-text content to draw conclusions relying upon the analysis of papers' semantic similarity.

\section{Conclusion}\label{sec:conclusion}

The identification of keywords from short texts poses a significant challenge. In this paper, we evaluated whether well-known approaches are able to extract keywords from abstracts that are compatible with a full-content analysis. 
Due to the limited context provided in abstracts, we employed methods that leverage the citation context to cluster papers into semantically similar groups. Additionally, we used strategies based on statistics and the K-Means algorithm. Reference keywords were obtained from the full content of papers through the use of multiple techniques.

The findings indicate that a simple approach such as the K-Means algorithm outperforms methods that rely on communities derived from citation networks. Additionally, the results demonstrate a similar performance among the various community detection methods applied to citation networks, with no clear superiority demonstrated by any particular method.

All in all, citation networks and alternative methods that do not rely on citations  demonstrated suboptimal performance. This result implies that the keywords obtained from abstracts are not consistent with those obtained from a comprehensive content analysis. Consequently, further research is necessary to investigate whether the observed variations may lead to discrepancies in the analysis of document similarity networks~\cite{minaee2021deep}.

One way to potentially enhance the performance of clustering methods for keyword extraction is through the use of alternative methods for text vector representation. The incorporation of text embeddings, such as those generated by the BERT model~\cite{beltagy2019scibert}, may assist in effectively representing the documents. Additionally, implementing synonym handling during the generation of reference keywords could also prove beneficial. The performance could also be improved by integrating citation and text-based information when creating paper networks.

\section*{Acknowledgments}
This study was financed in part by the Coordenação de Aperfeiçoamento de Pessoal de Nível Superior - Brasil (CAPES) - Finance Code 001. Diego R. Amancio acknowledges financial support from CNPq-Brazil (Grant no. 304026/2018-2, 311074/2021-9)  and FAPESP (Grant no. 20/06271-0). Thiago C. Silva (Grant no. 308171/2019-5, 408546/2018-2) gratefully acknowledges financial support from the CNPq foundation.

\section*{Appendix}

\subsection{Community detection methods}\label{app:community}

In this section, we provide a brief description of the network community methods employed in this paper:

\begin{enumerate}

    \item \textit{Multilevel}:  in this algorithm, each node is assigned to a different community. Then  nodes are moved to the communities of their corresponding neighbors that yields the highest positive contribution to modularity~\cite{Blondel_2008}. This process is repeated until the local contribution of nodes to the modularity is no longer improved. 
    Each community from the original network is reduced into a single node (maintaining the total weight of the adjacent edges) and the method continues to the next level. The algorithm ends when there is no longer any possibility of increasing the modularity score after reducing communities to nodes.
    
    \item \textit{Label propagation:} The method presented in~\cite{PhysRevE.76.036106} is based on the principles of neighborhood connectivity and information diffusion in networks. The approach begins by assigning unique community labels to each node in the network. These labels are subsequently propagated throughout the network. During each iteration, each node adopts the most prevalent label within its immediate neighborhood. The edges within the network are then randomly removed, and the nodes are updated in a random order before the next iteration commences. The algorithm stops when the nodes reach a consensus, which is defined as a state in which each node holds the majority label among its neighboring nodes.
    
    \item \textit{Leiden:} The Leiden algorithm, which was proposed in~\cite{traag2019louvain}, represents an improvement to the widely-used multilevel method~\cite{Blondel_2008}. The latter is known to have a weakness of often discovering communities that are weakly connected. In contrast, the Leiden method aims at ensuring that communities are well-connected through the implementation of the following three phases:
    (i) local moving of nodes (as in the multilevel method); (ii) refinement of partitions; and (iii) aggregation of the network. By incorporating these three phases, the Leiden algorithm is able to uncover higher-quality clusters in significantly less time when compared to the multilevel method.
    
    \item \textit{Fast Greedy:} this algorithm is based on hierarchical agglomerative clustering and aims to optimize the modularity score~\cite{clauset2004finding}. The method begins by considering a subnetwork composed exclusively of edges between highly-connected nodes. This methodology subsequently evaluates randomly selected edges that improve the modularity of the subnetwork and aggregates them. This process is repeated until the incremental improvement in modularity becomes negligible. Finally, the communities are obtained by identifying the connected components within the subnetwork.
    
    \item \textit{Infomap:} the algorithm  was introduced by~\cite{rosvall2008maps} and is based on information theory. This method begins by encoding the network into modules in a manner that maximizes the amount of information retained from the original network. The encoded network is then transmitted through a channel with limited capacity. The goal of the decoder is to attempt to decode the message and construct a set of possible candidates for the original network. The fewer the number of candidates, the more information about the original network has been transmitted.  The algorithm also uses random walks to analyze the flow of information through the network.
    
\end{enumerate}

\subsection{Statistical keyword detection}\label{app:statistical}

\begin{itemize}

        \item \textit{Word frequency and tf-idf methods (Freq. and tf-idf):} one of the simplest techniques for keyword extraction is the frequency-based approach, which assigns relevance to words that occur at a high frequency. The words that rank the highest in terms of frequency are therefore considered as keywords. In order to mitigate the limitations of the frequency-based methods, we also evaluated the tf-idf method. Unlike the frequency-based approach, the tf-idf method assigns a weight to the frequency of each word based on its number of occurrences within the document as well as throughout the entire dataset. In this approach, the words with the highest tf-idf values are considered as keywords.  
        
        \item \textit{Word entropy (W.E.):} This method leverages Shannon's entropy to analyze the information content of the sequence of occurrences of each word in a given text~\cite{carretero2013improving}. This technique requires partitioning the texts into $N$ segments to calculate the entropy of each word. In this study, we partitioned the paper texts according to the number of sentences that make up each text. According to this method, the higher the value of entropy of a word, the greater the heterogeneity of the distribution of that word within the text, and thus the greater its relevance. One of the key advantages of this method is that it does not require a large text corpus for training; it only requires the input text.
         
        \item \textit{Word intermittency (W.I.):} This metric takes into account the relationship between the significance of a word and its spatial distribution~\cite{amancio2013probing}. Previous research has found that important words are closely related to the main topics of the text and display a highly heterogeneous distribution. Such words tend to be located in specific regions of the text, exhibit large frequency fluctuations and often form clusters~\cite{carretero2013improving}. In contrast, common words such as stopwords are distributed randomly throughout the document and exhibit a relatively homogeneous distribution. Thus, as proposed by ~\cite{carretero2013improving}, a statistical analysis of the distribution of word occurrences can be employed to identify relevant keywords within a given text. Similar to the frequency and entropy methods, this technique identifies important words solely based on the target text and does not require external information. 
        
        \item \textit{Yake}: The Yake method extracts statistical features from the source text to identify the most relevant keywords~\cite{campos2020yake}. Five features are computed for each individual term: (i) casing, (ii) word positional, which assigns greater importance to words that appear at the beginning of a text, (iii) word frequency, which assigns relevance to words that occur more frequently, (iv) word relatedness to context, which measures the number of different terms that appear to the left and right of the target word, and (v) word \emph{difSentence}, which measures how often a word appears across different sentences. These features are then combined into a single measure to assign an importance weight to each word. According to this score, words with the lower values are considered as relevant keywords~\cite{campos2018yake}.

        \item \textit{TextRank (TextR):} The TextRank method, proposed by~\cite{mihalcea2004textrank}, is a graph-based approach that employs the PageRank algorithm and is widely used for text summarization and keyword extraction tasks. In this method, texts are modeled as word co-occurrence networks, where the nodes are represented by words and edges are established between two nodes if they co-occur within a window size. In the original paper, the window size was set between 2 and 10 words. The PageRank algorithm is employed to rank each word, and the top-ranked words are selected as relevant keywords.
        
        \item \textit{BERT-based method}: the Bidirectional Encoder Representations from Transformers (BERT) technique is a state-of-the-art embedding model that captures the semantic content of documents through dense vector representations~\cite{devlin2018bert}. The BERT-based method generates word embeddings for each n-gram in the text. Subsequently, the cosine similarity metric is applied to identify the words that are most similar to the original document. The top-ranking similar words are then considered as relevant keywords for each document.
        
        \item \textit{Voting system (V.S.):} In order to improve the accuracy of the long text keyword extractor, we combined the results of the above proposed methods. We used a voting system based on the keywords found by most keyword extraction methods.
    \end{itemize}

\subsection{Complete results based on accuracy analysis}\label{app:results}

\begin{table}[h!]
\caption{Accuracy obtained from the evaluation of keyword extraction methods for short texts (paper abstracts). Here we considered as reference keywords the relevant words found by the \emph{traditional and statistical} methods for the full content of the papers. $N$ represents the number of top keywords we recovered for both short and long texts keyword extraction methods.}
\label{tab:accuracy_statistical}
\begin{tabular}{c|cccc|cccc|}
\cline{2-9} & \multicolumn{4}{c|}{\textbf{Word Frequency}} & \multicolumn{4}{c|}{\textbf{tf-idf}} \\ \hline
\multicolumn{1}{|c|}{\textbf{method}} & \multicolumn{1}{c}{\textbf{N=10}} & \multicolumn{1}{c}{\textbf{N=20}} & \multicolumn{1}{c}{\textbf{N=30}} & \textbf{N=40} & \multicolumn{1}{c}{\textbf{N=10}} & \multicolumn{1}{c}{\textbf{N=20}} & \multicolumn{1}{c}{\textbf{N=30}} & \textbf{N=40} \\ \hline

\multicolumn{1}{|c|}{multilevel}      & \multicolumn{1}{c}{0.1578}        & \multicolumn{1}{c}{0.2559}        & \multicolumn{1}{c}{0.3269}        & 0.3678        & \multicolumn{1}{c}{0.0740}        & \multicolumn{1}{c}{0.1066}        & \multicolumn{1}{c}{0.1448}        & 0.1793        \\  

\multicolumn{1}{|c|}{labelProp}       & \multicolumn{1}{c}{0.1375}              & \multicolumn{1}{c}{0.2312}              & \multicolumn{1}{c}{0.3022}              & 0.3504              & \multicolumn{1}{c}{0.0787}              & \multicolumn{1}{c}{0.1033}              & \multicolumn{1}{c}{0.1363}              &  0.1724             \\  

\multicolumn{1}{|c|}{leiden}          & \multicolumn{1}{c}{0.1496}              & \multicolumn{1}{c}{0.2533}              & \multicolumn{1}{c}{0.3260}              & 0.3664               & \multicolumn{1}{c}{0.0819}              & \multicolumn{1}{c}{0.1153}              & \multicolumn{1}{c}{0.1463}              &     0.1785          \\  

\multicolumn{1}{|c|}{fastG}           & \multicolumn{1}{c}{0.1465}              & \multicolumn{1}{c}{0.2518}              & \multicolumn{1}{c}{0.3290}              &  0.3684             & \multicolumn{1}{c}{0.0713}              & \multicolumn{1}{c}{0.1023}              & \multicolumn{1}{c}{0.1419}              &     0.1767          \\  

\multicolumn{1}{|c|}{infomap}         & \multicolumn{1}{c}{0.1375}              & \multicolumn{1}{c}{0.2320}              & \multicolumn{1}{c}{0.3175}              &   0.3647            & \multicolumn{1}{c}{0.1458}              & \multicolumn{1}{c}{0.1561}              & \multicolumn{1}{c}{0.1726}              &   0.1951            \\  

\multicolumn{1}{|c|}{tf-idf}           & \multicolumn{1}{c}{0.2858}              & \multicolumn{1}{c}{0.2889}              & \multicolumn{1}{c}{0.3071}              &     0.3350          & \multicolumn{1}{c}{0.3596}              & \multicolumn{1}{c}{0.3223}              & \multicolumn{1}{c}{0.2984}              &    0.2755           \\  

\multicolumn{1}{|c|}{KMeans}          & \multicolumn{1}{c}{0.4253}              & \multicolumn{1}{c}{0.4213}              & \multicolumn{1}{c}{0.4220}              &    0.4146           & \multicolumn{1}{c}{0.2129}              & \multicolumn{1}{c}{0.2106}              & \multicolumn{1}{c}{0.2198}              &    0.2279           \\ \hline

 & \multicolumn{4}{c|}{\textbf{Word Entropy}} & \multicolumn{4}{c|}{\textbf{Word Intermittency}} \\ \hline
 \multicolumn{1}{|c|}{\textbf{method}} & \multicolumn{1}{c}{\textbf{N=10}} & \multicolumn{1}{c}{\textbf{N=20}} & \multicolumn{1}{c}{\textbf{N=30}} & \textbf{N=40} & \multicolumn{1}{c}{\textbf{N=10}} & \multicolumn{1}{c}{\textbf{N=20}} & \multicolumn{1}{c}{\textbf{N=30}} & \textbf{N=40} \\ \hline
 
\multicolumn{1}{|c|}{multilevel}      & \multicolumn{1}{c}{0.1341}              & \multicolumn{1}{c}{0.2294}              & \multicolumn{1}{c}{0.3004}              &  0.3460             & \multicolumn{1}{c}{0.0679}              & \multicolumn{1}{c}{0.1291}              & \multicolumn{1}{c}{0.1828}              &   0.2233            \\

\multicolumn{1}{|c|}{labelProp}       & \multicolumn{1}{c}{0.1165}              & \multicolumn{1}{c}{0.2052}              & \multicolumn{1}{c}{0.2769}              &  0.3285             & \multicolumn{1}{c}{0.0608}              & \multicolumn{1}{c}{0.1159}              & \multicolumn{1}{c}{0.1663}              &     0.2110          \\

\multicolumn{1}{|c|}{leiden}          & \multicolumn{1}{c}{0.1279}              & \multicolumn{1}{c}{0.2293}              & \multicolumn{1}{c}{0.3007}              &    0.3455           & \multicolumn{1}{c}{0.0651}              & \multicolumn{1}{c}{0.1288}              & \multicolumn{1}{c}{0.1829}              &      0.2228         \\

\multicolumn{1}{|c|}{fastG}           & \multicolumn{1}{c}{0.1237}              & \multicolumn{1}{c}{0.2232}              & \multicolumn{1}{c}{0.3012}              &   0.3456            & \multicolumn{1}{c}{0.0658}              & \multicolumn{1}{c}{0.1241}              & \multicolumn{1}{c}{0.1807}              &  0.2218             \\

\multicolumn{1}{|c|}{infomap}         & \multicolumn{1}{c}{0.1259}              & \multicolumn{1}{c}{0.2166}              & \multicolumn{1}{c}{0.2967}              &  0.3441             & \multicolumn{1}{c}{0.0715}              & \multicolumn{1}{c}{0.1274}              & \multicolumn{1}{c}{0.1811}              &      0.2225         \\

\multicolumn{1}{|c|}{tf-idf}           & \multicolumn{1}{c}{0.2477}              & \multicolumn{1}{c}{0.2708}              & \multicolumn{1}{c}{0.2945}              &  0.3217             & \multicolumn{1}{c}{0.1093}              & \multicolumn{1}{c}{0.1539}              & \multicolumn{1}{c}{0.1868}              &     0.2180          \\

\multicolumn{1}{|c|}{KMeans}          & \multicolumn{1}{c}{0.3596}              & \multicolumn{1}{c}{0.3835}              & \multicolumn{1}{c}{0.3929}              & 0.3925
              & \multicolumn{1}{c}{0.1335}              & \multicolumn{1}{c}{0.1921}              & \multicolumn{1}{c}{0.2280}              &   0.2499            \\ \hline

 & \multicolumn{4}{c|}{\textbf{Yake}} & \multicolumn{4}{c|}{\textbf{TextRank}}  \\ \hline
\multicolumn{1}{|c|}{\textbf{method}} & \multicolumn{1}{c}{\textbf{N=10}} & \multicolumn{1}{c}{\textbf{N=20}} & \multicolumn{1}{c}{\textbf{N=30}} & \textbf{N=40} & \multicolumn{1}{c}{\textbf{N=10}} & \multicolumn{1}{c}{\textbf{N=20}} & \multicolumn{1}{c}{\textbf{N=30}} & \textbf{N=40} \\ \hline

\multicolumn{1}{|c|}{multilevel}      & \multicolumn{1}{c}{0.1553}              & \multicolumn{1}{c}{0.2536}              & \multicolumn{1}{c}{0.3242}              &  0.3650             & \multicolumn{1}{c}{0.1138}              & \multicolumn{1}{c}{0.1689}              & \multicolumn{1}{c}{0.2101}              &     0.2326          \\  

\multicolumn{1}{|c|}{labelProp}       & \multicolumn{1}{c}{0.1362}              & \multicolumn{1}{c}{0.2290}              & \multicolumn{1}{c}{0.2997}              &     0.3481          & \multicolumn{1}{c}{0.0967}              & \multicolumn{1}{c}{0.1594}              & \multicolumn{1}{c}{0.2012}              &  0.2260             \\  

\multicolumn{1}{|c|}{leiden}          & \multicolumn{1}{c}{0.1509}              & \multicolumn{1}{c}{0.2516}              & \multicolumn{1}{c}{0.3226}              &   0.3633            & \multicolumn{1}{c}{0.0986}              & \multicolumn{1}{c}{0.1673}              & \multicolumn{1}{c}{0.2090}              &       0.2319        \\  

\multicolumn{1}{|c|}{fastG}           & \multicolumn{1}{c}{0.1448}              & \multicolumn{1}{c}{0.2485}              & \multicolumn{1}{c}{0.3256}              &    0.3653           & \multicolumn{1}{c}{0.1172}              & \multicolumn{1}{c}{0.1818}              & \multicolumn{1}{c}{0.2184}              &      0.2356         \\  

\multicolumn{1}{|c|}{infomap}         & \multicolumn{1}{c}{0.1400}              & \multicolumn{1}{c}{0.2309}              & \multicolumn{1}{c}{0.3153}              &      0.3616         & \multicolumn{1}{c}{0.0829}              & \multicolumn{1}{c}{0.1503}              & \multicolumn{1}{c}{0.2034}              &      0.2302         \\  

\multicolumn{1}{|c|}{tf-idf}           & \multicolumn{1}{c}{0.2905}              & \multicolumn{1}{c}{0.2937}              & \multicolumn{1}{c}{0.3091}              &   0.3353            & \multicolumn{1}{c}{0.1381}              & \multicolumn{1}{c}{0.1492}              & \multicolumn{1}{c}{0.1703}              &     0.1951          \\  

\multicolumn{1}{|c|}{KMeans}          & \multicolumn{1}{c}{0.4221}              & \multicolumn{1}{c}{0.4190}              & \multicolumn{1}{c}{0.4194}              &    0.4121           & \multicolumn{1}{c}{0.2132}              & \multicolumn{1}{c}{0.2337}              & \multicolumn{1}{c}{0.2459}              &   0.2479            \\ \hline

 & \multicolumn{4}{c|}{\textbf{BERT}} & \multicolumn{4}{c|}{\textbf{Voting System}}\\ \hline
\multicolumn{1}{|c|}{\textbf{method}} & \multicolumn{1}{c}{\textbf{N=10}} & \multicolumn{1}{c}{\textbf{N=20}} & \multicolumn{1}{c}{\textbf{N=30}} & \textbf{N=40} & \multicolumn{1}{c}{\textbf{N=10}} & \multicolumn{1}{c}{\textbf{N=20}} & \multicolumn{1}{c}{\textbf{N=30}} & \textbf{N=40} \\ \hline

\multicolumn{1}{|c|}{multilevel}      & \multicolumn{1}{c}{0.0573}              & \multicolumn{1}{c}{0.0924}              & \multicolumn{1}{c}{0.1254}              & 0.1501              & \multicolumn{1}{c}{0.1231}              & \multicolumn{1}{c}{0.1925}              & \multicolumn{1}{c}{0.2688}              &  0.3283             \\  

\multicolumn{1}{|c|}{labelProp}       & \multicolumn{1}{c}{0.0710}              & \multicolumn{1}{c}{0.0936}              & \multicolumn{1}{c}{0.1221}              &  0.1462             & \multicolumn{1}{c}{0.1108}              & \multicolumn{1}{c}{0.1758}              & \multicolumn{1}{c}{0.2485}              &      0.3123         \\  

\multicolumn{1}{|c|}{leiden}          & \multicolumn{1}{c}{0.0757}              & \multicolumn{1}{c}{0.1092}              & \multicolumn{1}{c}{0.1325}              &    0.1518           & \multicolumn{1}{c}{0.1325}              & \multicolumn{1}{c}{0.2003}              & \multicolumn{1}{c}{0.2716}              &     0.3274          \\  

\multicolumn{1}{|c|}{fastG}           & \multicolumn{1}{c}{0.0739}              & \multicolumn{1}{c}{0.0965}              & \multicolumn{1}{c}{0.1257}              &   0.1494            & \multicolumn{1}{c}{0.1300}              & \multicolumn{1}{c}{0.1941}              & \multicolumn{1}{c}{0.2702}              &       0.3280        \\  

\multicolumn{1}{|c|}{infomap}         & \multicolumn{1}{c}{0.0596}              & \multicolumn{1}{c}{0.0938}              & \multicolumn{1}{c}{0.1261}              &    0.1483           & \multicolumn{1}{c}{0.1115}              & \multicolumn{1}{c}{0.1826}              & \multicolumn{1}{c}{0.2633}              &     0.3251          \\  

\multicolumn{1}{|c|}{tf-idf}           & \multicolumn{1}{c}{0.0899}              & \multicolumn{1}{c}{0.1043}              & \multicolumn{1}{c}{0.1207}              &  0.1394             & \multicolumn{1}{c}{0.1891}              & \multicolumn{1}{c}{0.2280}              & \multicolumn{1}{c}{0.2687}              &      0.3070         \\  

\multicolumn{1}{|c|}{KMeans}          & \multicolumn{1}{c}{0.1804}              & \multicolumn{1}{c}{0.1668}              & \multicolumn{1}{c}{0.1688}              &  0.1726             & \multicolumn{1}{c}{0.2761}              & \multicolumn{1}{c}{0.3110}              & \multicolumn{1}{c}{0.3531}              &   0.3746            \\ \hline

\end{tabular}
\end{table}

\begin{table}[h!]
\caption{Accuracy obtained from the evaluation of keyword extraction methods for short texts (paper abstracts). Here we considered as reference keywords the most important words found by the \emph{network-based} methods for the full content of the papers. $N$ is the number of top keywords we recovered for both short and long texts keyword extraction methods.}
\label{tab:accuracy_network}
\begin{tabular}{c|cccc|cccc|}
\cline{2-9} & \multicolumn{4}{c|}{\textbf{Degree}} & \multicolumn{4}{c|}{\textbf{PageRank}} \\ \hline
\multicolumn{1}{|c|}{\textbf{method}} & \multicolumn{1}{c}{\textbf{N=10}} & \multicolumn{1}{c}{\textbf{N=20}} & \multicolumn{1}{c}{\textbf{N=30}} & \textbf{N=40} & \multicolumn{1}{c}{\textbf{N=10}} & \multicolumn{1}{c}{\textbf{N=20}} & \multicolumn{1}{c}{\textbf{N=30}} & \textbf{N=40} \\ \hline

\multicolumn{1}{|c|}{multilevel}      & \multicolumn{1}{c}{0.1605}        & \multicolumn{1}{c}{0.2490}        & \multicolumn{1}{c}{0.3149}        & 0.3539        & \multicolumn{1}{c}{0.1622}        & \multicolumn{1}{c}{0.2498}        & \multicolumn{1}{c}{0.3153}        & 0.3514        \\  

\multicolumn{1}{|c|}{labelProp}       & \multicolumn{1}{c}{0.1324}              & \multicolumn{1}{c}{0.2234}              & \multicolumn{1}{c}{0.2903}              &    0.3374           & \multicolumn{1}{c}{0.1329}              & \multicolumn{1}{c}{0.2247}              & \multicolumn{1}{c}{0.2899}              &       0.3349        \\  

\multicolumn{1}{|c|}{leiden}          & \multicolumn{1}{c}{0.1417}              & \multicolumn{1}{c}{0.2421}              & \multicolumn{1}{c}{0.3125}              &     0.3522          & \multicolumn{1}{c}{0.1428}              & \multicolumn{1}{c}{0.2429}              & \multicolumn{1}{c}{0.3120}              &     0.3494          \\  

\multicolumn{1}{|c|}{fastG}           & \multicolumn{1}{c}{0.1466}              & \multicolumn{1}{c}{0.2496}              & \multicolumn{1}{c}{0.3195}              &   0.3549            & \multicolumn{1}{c}{0.1481}              & \multicolumn{1}{c}{0.2510}              & \multicolumn{1}{c}{0.3194}              &    0.3523           \\  

\multicolumn{1}{|c|}{infomap}         & \multicolumn{1}{c}{0.1256}              & \multicolumn{1}{c}{0.2152}              & \multicolumn{1}{c}{0.3002}              &   0.3476            & \multicolumn{1}{c}{0.1247}              & \multicolumn{1}{c}{0.2146}              & \multicolumn{1}{c}{0.2979}              &    0.3445           \\  

\multicolumn{1}{|c|}{tf-idf}           & \multicolumn{1}{c}{0.2529}              & \multicolumn{1}{c}{0.2546}              & \multicolumn{1}{c}{0.2757}              &  0.3085             & \multicolumn{1}{c}{0.2516}              & \multicolumn{1}{c}{0.2521}              & \multicolumn{1}{c}{0.2725}              &        0.3043       \\  

\multicolumn{1}{|c|}{KMeans}          & \multicolumn{1}{c}{0.4047}              & \multicolumn{1}{c}{0.3989}              & \multicolumn{1}{c}{0.4010}              & 0.3948              & \multicolumn{1}{c}{0.4070}              & \multicolumn{1}{c}{0.3998}              & \multicolumn{1}{c}{0.3999}              &     0.3920          \\ \hline

 & \multicolumn{4}{c|}{\textbf{Betweenness}} & \multicolumn{4}{c|}{\textbf{Eigenvector}} \\ \hline
 \multicolumn{1}{|c|}{\textbf{method}} & \multicolumn{1}{c}{\textbf{N=10}} & \multicolumn{1}{c}{\textbf{N=20}} & \multicolumn{1}{c}{\textbf{N=30}} & \textbf{N=40} & \multicolumn{1}{c}{\textbf{N=10}} & \multicolumn{1}{c}{\textbf{N=20}} & \multicolumn{1}{c}{\textbf{N=30}} & \textbf{N=40} \\ \hline
 
\multicolumn{1}{|c|}{multilevel}      & \multicolumn{1}{c}{0.1563}              & \multicolumn{1}{c}{0.2281}              & \multicolumn{1}{c}{0.2809}              &  0.3127             & \multicolumn{1}{c}{0.1484}              & \multicolumn{1}{c}{0.2385}              & \multicolumn{1}{c}{0.3008}              &      0.3385         \\

\multicolumn{1}{|c|}{labelProp}       & \multicolumn{1}{c}{0.1255}              & \multicolumn{1}{c}{0.2021}              & \multicolumn{1}{c}{0.2571}              &    0.2973           & \multicolumn{1}{c}{0.1253}              & \multicolumn{1}{c}{0.2136}              & \multicolumn{1}{c}{0.2787}              &   0.3238            \\

\multicolumn{1}{|c|}{leiden}          & \multicolumn{1}{c}{0.1356}              & \multicolumn{1}{c}{0.2202}              & \multicolumn{1}{c}{0.2774}              &   0.3099            & \multicolumn{1}{c}{0.1343}              & \multicolumn{1}{c}{0.2315}              & \multicolumn{1}{c}{0.2994}              &        0.3370       \\

\multicolumn{1}{|c|}{fastG}           & \multicolumn{1}{c}{0.1392}              & \multicolumn{1}{c}{0.2272}              & \multicolumn{1}{c}{0.2836}              &   0.3134             & \multicolumn{1}{c}{0.1387}              & \multicolumn{1}{c}{0.2373}              & \multicolumn{1}{c}{0.3031}              &       0.3393        \\

\multicolumn{1}{|c|}{infomap}         & \multicolumn{1}{c}{0.1125}              & \multicolumn{1}{c}{0.1855}              & \multicolumn{1}{c}{0.2596}              &   0.3028            & \multicolumn{1}{c}{0.1239}              & \multicolumn{1}{c}{0.2120}              & \multicolumn{1}{c}{0.2921}              &    0.3346           \\

\multicolumn{1}{|c|}{tf-idf}           & \multicolumn{1}{c}{0.2291}              & \multicolumn{1}{c}{0.2255}              & \multicolumn{1}{c}{0.2387}              &   0.2661            & \multicolumn{1}{c}{0.2490}              & \multicolumn{1}{c}{0.2522}              & \multicolumn{1}{c}{0.2726}              &    0.3021           \\

\multicolumn{1}{|c|}{KMeans}          & \multicolumn{1}{c}{0.3828}              & \multicolumn{1}{c}{0.3656}              & \multicolumn{1}{c}{0.3574}              &     0.3503          & \multicolumn{1}{c}{0.3865}              & \multicolumn{1}{c}{0.3844}              & \multicolumn{1}{c}{0.3847}              &      0.3797         \\ \hline

 & \multicolumn{4}{c|}{\textbf{Closeness}} & \multicolumn{4}{c|}{\textbf{Accessibility (h=1)}}  \\ \hline
\multicolumn{1}{|c|}{\textbf{method}} & \multicolumn{1}{c}{\textbf{N=10}} & \multicolumn{1}{c}{\textbf{N=20}} & \multicolumn{1}{c}{\textbf{N=30}} & \textbf{N=40} & \multicolumn{1}{c}{\textbf{N=10}} & \multicolumn{1}{c}{\textbf{N=20}} & \multicolumn{1}{c}{\textbf{N=30}} & \textbf{N=40} \\ \hline

\multicolumn{1}{|c|}{multilevel}      & \multicolumn{1}{c}{0.1585}              & \multicolumn{1}{c}{0.2438}              & \multicolumn{1}{c}{0.3066}              &      0.3426         & \multicolumn{1}{c}{0.1605}              & \multicolumn{1}{c}{0.2490}              & \multicolumn{1}{c}{0.3149}              &       0.3539        \\  

\multicolumn{1}{|c|}{labelProp}       & \multicolumn{1}{c}{0.1317}              & \multicolumn{1}{c}{0.2198}              & \multicolumn{1}{c}{0.2855}              &      0.3278         & \multicolumn{1}{c}{0.1324}              & \multicolumn{1}{c}{0.2234}              & \multicolumn{1}{c}{0.2903}              &      0.3374         \\  

\multicolumn{1}{|c|}{leiden}          & \multicolumn{1}{c}{0.1390}              & \multicolumn{1}{c}{0.2373}              & \multicolumn{1}{c}{0.3052}              &   0.3411            & \multicolumn{1}{c}{0.1417}              & \multicolumn{1}{c}{0.2421}              & \multicolumn{1}{c}{0.3125}              &      0.3522         \\  

\multicolumn{1}{|c|}{fastG}           & \multicolumn{1}{c}{0.1461}              & \multicolumn{1}{c}{0.2446}              & \multicolumn{1}{c}{0.3096}              &   0.3431            & \multicolumn{1}{c}{0.1466}              & \multicolumn{1}{c}{0.2496}              & \multicolumn{1}{c}{0.3195}              &     0.3549          \\  

\multicolumn{1}{|c|}{infomap}         & \multicolumn{1}{c}{0.1209}              & \multicolumn{1}{c}{0.2127}              & \multicolumn{1}{c}{0.2932}              &     0.3367          & \multicolumn{1}{c}{0.1256}              & \multicolumn{1}{c}{0.2152}              & \multicolumn{1}{c}{0.3002}              &  0.3476             \\  

\multicolumn{1}{|c|}{tf-idf}           & \multicolumn{1}{c}{0.2478}              & \multicolumn{1}{c}{0.2499}              & \multicolumn{1}{c}{0.2717}              &   0.3019            & \multicolumn{1}{c}{0.2529}              & \multicolumn{1}{c}{0.2546}              & \multicolumn{1}{c}{0.2757}              &        0.3085       \\  

\multicolumn{1}{|c|}{KMeans}          & \multicolumn{1}{c}{0.3936}              & \multicolumn{1}{c}{0.3883}              & \multicolumn{1}{c}{0.3891 }              &    0.3828           & \multicolumn{1}{c}{0.4047}              & \multicolumn{1}{c}{0.3989}              & \multicolumn{1}{c}{0.4010}              &         0.3948      \\ \hline

 & \multicolumn{4}{c|}{\textbf{Accessibility(h=2)}} & \multicolumn{4}{c|}{\textbf{Voting System}}\\ \hline
\multicolumn{1}{|c|}{\textbf{method}} & \multicolumn{1}{c}{\textbf{N=10}} & \multicolumn{1}{c}{\textbf{N=20}} & \multicolumn{1}{c}{\textbf{N=30}} & \textbf{N=40} & \multicolumn{1}{c}{\textbf{N=10}} & \multicolumn{1}{c}{\textbf{N=20}} & \multicolumn{1}{c}{\textbf{N=30}} & \textbf{N=40} \\ \hline

\multicolumn{1}{|c|}{multilevel}      & \multicolumn{1}{c}{0.1196}              & \multicolumn{1}{c}{0.1943}              & \multicolumn{1}{c}{0.2526}              &     0.2904          & \multicolumn{1}{c}{0.1605}              & \multicolumn{1}{c}{0.2481}              & \multicolumn{1}{c}{0.3141}              &     0.3535          \\  

\multicolumn{1}{|c|}{labelProp}       & \multicolumn{1}{c}{0.1048}              & \multicolumn{1}{c}{0.1744}              & \multicolumn{1}{c}{0.2338}              &   0.2777            & \multicolumn{1}{c}{0.1324}              & \multicolumn{1}{c}{0.2226}              & \multicolumn{1}{c}{0.2909}              &      0.3377         \\  

\multicolumn{1}{|c|}{leiden}          & \multicolumn{1}{c}{0.1088}              & \multicolumn{1}{c}{0.1869}              & \multicolumn{1}{c}{0.2508}              &      0.2894         & \multicolumn{1}{c}{0.1426}              & \multicolumn{1}{c}{0.2408}              & \multicolumn{1}{c}{0.3119}              &       0.3522        \\  

\multicolumn{1}{|c|}{fastG}           & \multicolumn{1}{c}{0.1108}              & \multicolumn{1}{c}{0.1909}              & \multicolumn{1}{c}{0.2519}              &  0.2899             & \multicolumn{1}{c}{0.1464}              & \multicolumn{1}{c}{0.2487}              & \multicolumn{1}{c}{0.3175}              &     0.3548          \\  

\multicolumn{1}{|c|}{infomap}         & \multicolumn{1}{c}{0.0997}              & \multicolumn{1}{c}{0.1740}              & \multicolumn{1}{c}{0.2419}              &   0.2855            & \multicolumn{1}{c}{0.1255}              & \multicolumn{1}{c}{0.2147}              & \multicolumn{1}{c}{0.3001}              &    0.3479           \\  

\multicolumn{1}{|c|}{tf-idf}           & \multicolumn{1}{c}{0.1977}              & \multicolumn{1}{c}{0.2125}              & \multicolumn{1}{c}{0.2344}              &     0.2638          & \multicolumn{1}{c}{0.2536}              & \multicolumn{1}{c}{0.2546}              & \multicolumn{1}{c}{0.2759}              &      0.3093         \\  

\multicolumn{1}{|c|}{KMeans}          & \multicolumn{1}{c}{0.3162}              & \multicolumn{1}{c}{0.3167}              & \multicolumn{1}{c}{0.3236}              &    0.3264           & \multicolumn{1}{c}{0.4039}              & \multicolumn{1}{c}{0.3963}              & \multicolumn{1}{c}{0.3985 }              &     0.3952          \\ \hline

\end{tabular}
\end{table}

\newpage

\bibliographystyle{ieeetr}
\bibliographystyle{abbrv}

\end{document}